\documentclass[aps,pra,showpacs,onecolumn,superscriptaddress]{revtex4}
\usepackage{amssymb}
\usepackage{graphicx}
\usepackage{epstopdf}
\usepackage{subfigure}
\usepackage{appendix}
\usepackage{dcolumn}
\usepackage{enumitem}
\usepackage{indentfirst}
\usepackage{bm}
\usepackage{amsmath}
\usepackage{mathtools}
\usepackage{epstopdf}
\usepackage{xcolor}
\usepackage{float}
\usepackage{braket}
\usepackage[english]{babel}
\usepackage{epstopdf}
\usepackage[colorlinks,linkcolor=magenta,citecolor=blue,urlcolor=blue]{hyperref}
\begin{document}




\title{Quantum criticality driven by the cavity coupling in the Rabi-dimer model}

\author{Shujie Cheng}
\affiliation{Department of Physics, Zhejiang Normal University, Jinhua 321004, China}
\author{He-Guang Xu}
\affiliation{School of Physics, Dalian University of Technology, Dalian 116024, China}
\author{Xueying Liu}
\affiliation{Department of Physics, Zhejiang Normal University, Jinhua 321004, China}
\author{Gao Xianlong }
\thanks{Corresponding author:~gaoxl@zjnu.edu.cn}
\affiliation{Department of Physics, Zhejiang Normal University, Jinhua 321004, China}

\date{\today}

\begin{abstract}
The superradiant phase transition (SPT) controlled by the interacting strength between the two-level atom
and the photons has been a hot topic in the Rabi model and the Rabi-dimer model. The
latter describes two Rabi cavities coupled with an inter-cavity hopping parameter. Moreover, the
SPT in the Rabi-dimer model is found to be the same universal class  that in the Rabi model by
investigating the correlation-length critical exponent. In this paper, we are concerned about whether the
inter-cavity hopping parameter between two Rabi cavities (i.e., the Rabi-dimer model) will induce
the SPT and to which the universal class of the phase transition belongs. We analytically derive the
phase boundary of the SPT and investigate the ground-state properties of the system. We uncover
that the inter-cavity induced SPT can be apparently understood from the ground-state energy and
the ground-state photon population, as well as the ground-state expectation value of the squared
anti-symmetric mode. From the scaling analysis of the fidelity susceptibility, we numerically verify
that the SPT driven by the cavity coupling belongs to the same universal class as the one driven by
the atom-cavity interaction. Our work enriches the studies on the SPT and its critical behaviors in
the Rabi-dimer model.
\end{abstract}

%\pacs{64.60.-i, 05.70.Fh, 64.60.Fr}
\maketitle

\section{Introduction}
The quantum system involving single-mode cavity interacting with a cloud of two-level atoms can be
described by the Dicke model \cite{Dicke}. When the interaction strength exceeds the critical value, the system will
undergo from a normal phase to a superradiant phase. This quantum behavior, is known as the superradiant phase
transition (SPT). During the decades, the investigation on the SPT has involved in many experimental mediums,
ranging from the atomic or molecular clouds \cite{am1,am2,am3,am4,am5,am6,am7}, nuclei \cite{nu1}, Rydberg atomic
gas \cite{Rg1,Rg2,Rg3,Rg4}, to superconducting qubits \cite{sq1}, and semiconductors \cite{se1,se2}. Recently, some
attentions from the perspective of theoretical schemes \cite{ts1,ts2} and experimental engineerings \cite{exp1,exp2,exp3,exp4,exp5}
are focused on how to employ the superradiant feature to design a new generation of extremely stable lasers.

In fact, beyond the multiatomic system, there appears superradiant phenomenon as well. A representative system
is the quantum Rabi model \cite{Rabi,book}, which describes a single two-level atom interacting with a single-mode cavity.
To investigate the SPT in the Rabi model, it is necessary to obtain the energy spectrum information and wavefunctions.
However, it has not been clear for a long time whether the Rabi model has analytical solutions of the eigenvalues
and wavefunctions. More are focused on the aspects of quasi-exact solutions \cite{am_1,am_3} and the analytically
explicit but approximate studies \cite{am_2,am_4,am_5,am_6,am_7,am_8}. However, Braak extracted the exact solution
of the Rabi model \cite{es1} in virtue of the Bargmann space \cite{bs} and  Chen et al. obtained the exact
solution of a two-photon Rabi model by using the Bogoliubov transformations \cite{es2}. Regarding the SPT, it is clear
that the superradiance appears when the ratio of the atomic frequency and the photonic frequency approaches
the thermodynamic limit \cite{ext_1,ext_2,ext_3,ext_4,ext_5,ext_6}, by which the universal dynamics of the Rabi model
is well studied \cite{HP}. Here, the the thermodynamic limit in these models refers to the frequency ratio tending to infinity. Recently, this finding has stimulated the researches on the SPT in other photonic systems \cite{ps1,ps2,R_d1,qc_1,qc_2,qc_3}.

Among the studies on the SPT, there exist interests in studying the quantum criticality of the above-mentioned systems.
Wei et al. investigated that in the Rabi model the fidelity susceptibility presents finite-frequency scaling behaviors near
the SPT point \cite{qc_1}, and the correlation-length critical exponent (CLCE) is calculated as $\nu=3/2$ \cite{HP,qc_1}.
The Rabi-dimer (RD) model is a general extension of the Rabi model in the multi-cavity system, which describes two Rabi
cavities coupled by an inter-cavity hopping strength \cite{R_d1}. For the RD system, more attention is focused on the SPT
induced by the atom-cavity interacting strength. Mao et al. found that the asymptotic behaviors of the expectation value
of canonical coordinate and the first-order derivative of entanglement entropy conform to the same scaling function, and
the extracted critical exponent agrees with that of the single Rabi model \cite{qc_2}. Even including the $\mathbf{A}^2$-term
in the RD model, the CLCE remains the same \cite{qc_3}. Interestingly, we notice that the CLCE of the Rabi
and RD models is the same as the one in the Dicke model \cite{Dicke_1,Dicke_2} and the Lipkin-Meshkov-Glick (LMG) model \cite{LMG_1,LMG_2,LMG_3}.
It means that the known phase transitions in the normal Rabi model, RD model, Dicke model, and the LMG model all belong
to the same universal class. However, in the RD model, the study on the relationship between the inter-cavity hopping strength
and the SPT is lacking. Accordingly, whether such an inter-cavity hopping-dominated SPT belongs to the same universal class
remains unknown.

To answer the above questions, we take the following strategies: First of all, we introduce the RD model
and investigate the SPT by analyzing how the ground-state (GS) energy and GS photon population, as well
as the squared anti-symmetric normal mode response to the inter-cavity hopping strength in Sec.~\ref{S2}.
Furthermore, we study the quantum criticality and extract the CLCE by analyzing the scaling behavior of the
fidelity susceptibility in Sec.~\ref{S3}. A summary is presented in Sec.~\ref{S4}.

\section{Rabi-dimer model and SPT}\label{S2}
The general Hamiltonian ($\hbar=1$) of the quantum RM model is written as
\begin{equation}\label{eq1}
H(J)/\omega=H_{0}+JH_1,
\end{equation}
where $H_0$ denotes the Hamiltonian of two equivalent cavities and each of the cavities is described by the Rabi model.
Here, the equivalent cavities refer to that the photonic frequency, atomic frequency, and photon-atom
coupling strength of two cavities are the same, equaling to $\omega$, $\Omega$, and $\lambda$, respectively.

By introducing the dimensionless parameters $\eta=\Omega/\omega$ and $g=2\lambda/\sqrt{\Omega\omega}$, $H_{0}$ reads
\begin{equation}
H_{0}=\sum_{i=L,R}a^\dag_{i} a_{i} + \frac{\eta}{2}\sigma^{z}_{i} - \frac{g\sqrt{\eta}}{2}\left(a_{i}+a^\dag_{i}\right)\sigma^{x}_{i},
\end{equation}
where $L$ and $R$ are the notations of two cavities, and $a^\dag$ and $a$ are bosonic operators.

$H_{1}$ is the driving Hamiltonian coupling the two cavities
with the inter-cavity hopping strength $J$, defined as
\begin{equation}
H_{1}=\left( a^\dag_{L}+a_{L}\right) \left(a^\dag_{R} + a_{R}\right).
\end{equation}

To obtain the phase boundary of the inter-cavity coupling driven SPT in the ground state,
the Hamiltonian in Eq.~(\ref{eq1}) is written as
\begin{equation}\label{re_eq1}
\mathcal{H}=\frac{H}{\Omega}\equiv \frac{1}{\eta}H_{0}+\frac{J}{\eta}H_{1}.
\end{equation}

Replacing the creation and annihilation operators by the renormalized canonical
coordinates (i.e., $x_{i}=(a^\dag_{i}+a_{i})/\sqrt{2\eta}$ and $p_{i}=i(a^\dag_{i}-a_{i})\sqrt{\eta/2}$), we have
\begin{equation}
\mathcal{H}=\frac{1}{2}\sum_{i=L,R}\left[x^2_{i}+\frac{p^2_i}{\eta^2}+\sigma^z_i-\sqrt{2}g\sigma^{x}_ix_i\right]+2Jx_{L}x_{R},
\end{equation}
and the symmetric ($x_{+}$) and anti-symmetric ($x_{-}$) normal modes satisfy $x_{\pm}=(x_{L}\pm x_{R})/\sqrt{2}$ \cite{qc_2}.

In the large-$\eta$ limit ($\eta\rightarrow\infty$), $\mathcal{H}$ further becomes
\begin{equation}
\mathcal{H}=\frac{1}{2}\sum_{i=L,R}\left[x^2_{i}+\sigma^{z}_{i}-\sqrt{2}g\sigma^{x}_{i}x_{i}\right]+2Jx_{L}x_{R},
\end{equation}
whose energies are
\begin{equation}
E_{\pm}(x_{L},x_{R})=\frac{1}{2}\sum_{i=L,R}\left(x^2_{i}\pm\sqrt{1+2g^2x^2_{i}}\right)+2Jx_{L}x_{R}.
\end{equation}

The low-energy branch $E_{-}(x_{L},x_{R})$ reflects the ground-state properties of the system. Truncating the expansion
of $E_{-}(x_{L},x_{R})$ to the quadratic term ($\mathcal{O}(x^4_{i})$), we have
\begin{equation}
E_{-}=\frac{1}{2}\psi^{T}\Lambda\psi,
\end{equation}
with $\Lambda=\left(\begin{array}{cc}1-g^2 & 2J \\ 2J & 1-g^2 \end{array}\right)$ and $\psi=(x_{L},x_{R})^{T}$.

\begin{figure}[H]
		\centering
		\includegraphics[width=\textwidth]{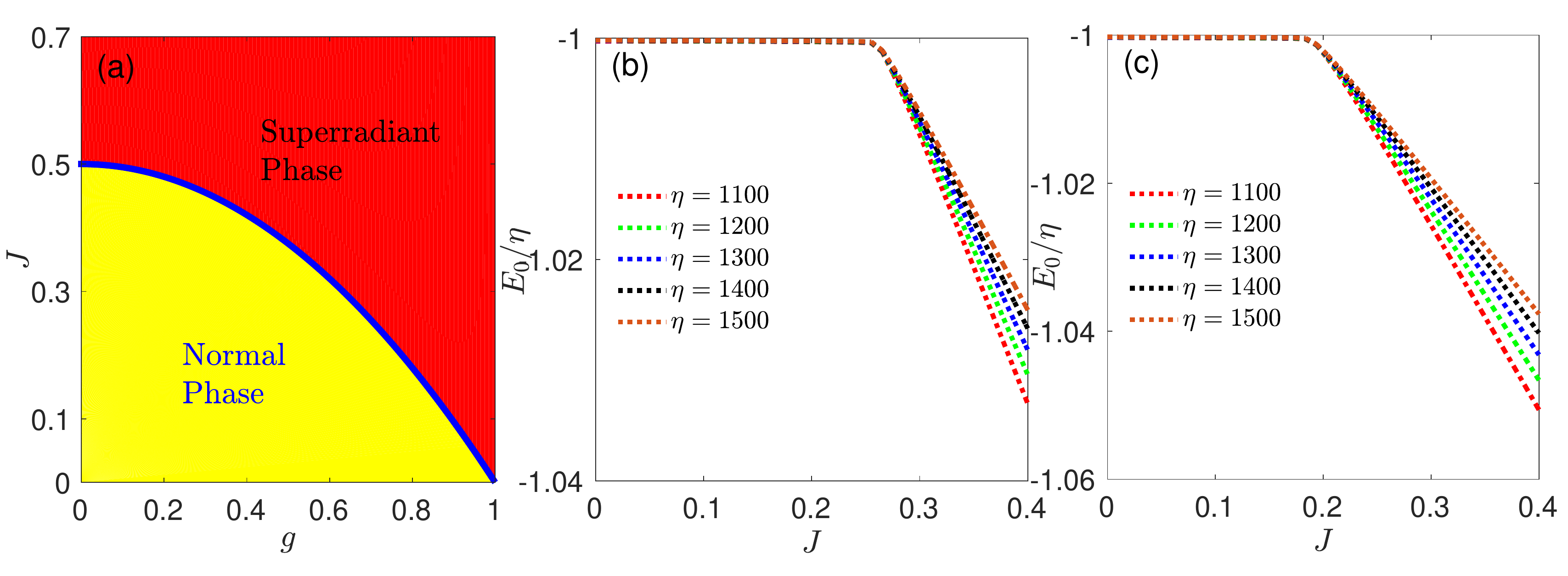}
		\caption{(Color Online) (a) Phase diagram in the $J$--$g$ parameter space. The blue solid line
		indicates the analytically obtained phase boundary ($J_{c}=(1-g^2)/2$), separating the normal
		phase (yellow region) from the superradiant one (red region). The GS energy $E_{0}$ as a function
		of the inter-cavity hopping strength $J$ with $g=0.7$ (b) and $g=0.8$ (c). The colors of the curves
		correspond to different $\eta$.}
		\label{f1}
\end{figure}

The eigenvalues $\lambda_{\pm}$ of the matrix $\Lambda$ are $\lambda_{\pm}=1-g^2\pm 2 J$. According to the Refs.~\cite{qc_2,qc_3},
the phase boundary can be extracted from $\lambda_{-}=0$. Accordingly, the critical point $J_{c}$ is obtained as $J_{c}=(1-g^2)/2$.
From the expression of $J_{c}$ and the phase diagram in Fig.~\ref{f1}(a), we intuitively know that the critical point of SPT gradually tends
to zero with the increase of $g$.

To further understand the properties of SPT, in the following, we investigate how the ground state energy, the number
of ground state photons, and the GS expectation value of the squared anti-symmetric normal mode response to the
cavity coupling. The Hamiltonian matrix of the RD model is constructed by the Fock states and the GS energy ($E_{0}$)
and the GS wave function is calculated by the Lanczos method. Without loss of generality, two typical cases with $g=0.7$
and $g=0.8$ are chosen for the numerical analysis. Other parameter spaces are tested supporting the SPT we found.
In appendix, we study the system with smaller $g=0.5$. In this case, the SPT phenomenon remains, and the quantum
criticality is the same as the ones of g=0.7 and 0.8 (see details in \ref{A}). At first, we present the numerical results
of the GS energy $E_{0}$ as the varying of the hopping strength $J$ for various $\eta=1100, 1200, 1300, 1400, 1500$
in Figs.~\ref{f1}(b) and \ref{f1}(c), with atom-cavity interacting strengths $g=0.7$ and $g=0.8$, respectively.
Intuitively, one can see an obvious phenomenon that when $J$ is small, $E_{0}$ is stable at $E_{0}\approx-\eta$,
and when $J$ exceeds a critical point, $E_{0}$ moves towards lower values. This abrupt transition behavior of the
GS energy implies the appearance of the SPT in this RD model. When $E_{0}\approx-\eta$, the system is in the normal
phase while it is superradiant when $E_{0}$ is much lower than $-\eta$. Besides, we notice that the transition point in
the $g=0.8$ case is smaller than that in the $g=0.7$ case. The stronger atom-cavity interaction will make the system superradiant.

\begin{figure}[htp]
		\centering
		\includegraphics[width=0.7\textwidth]{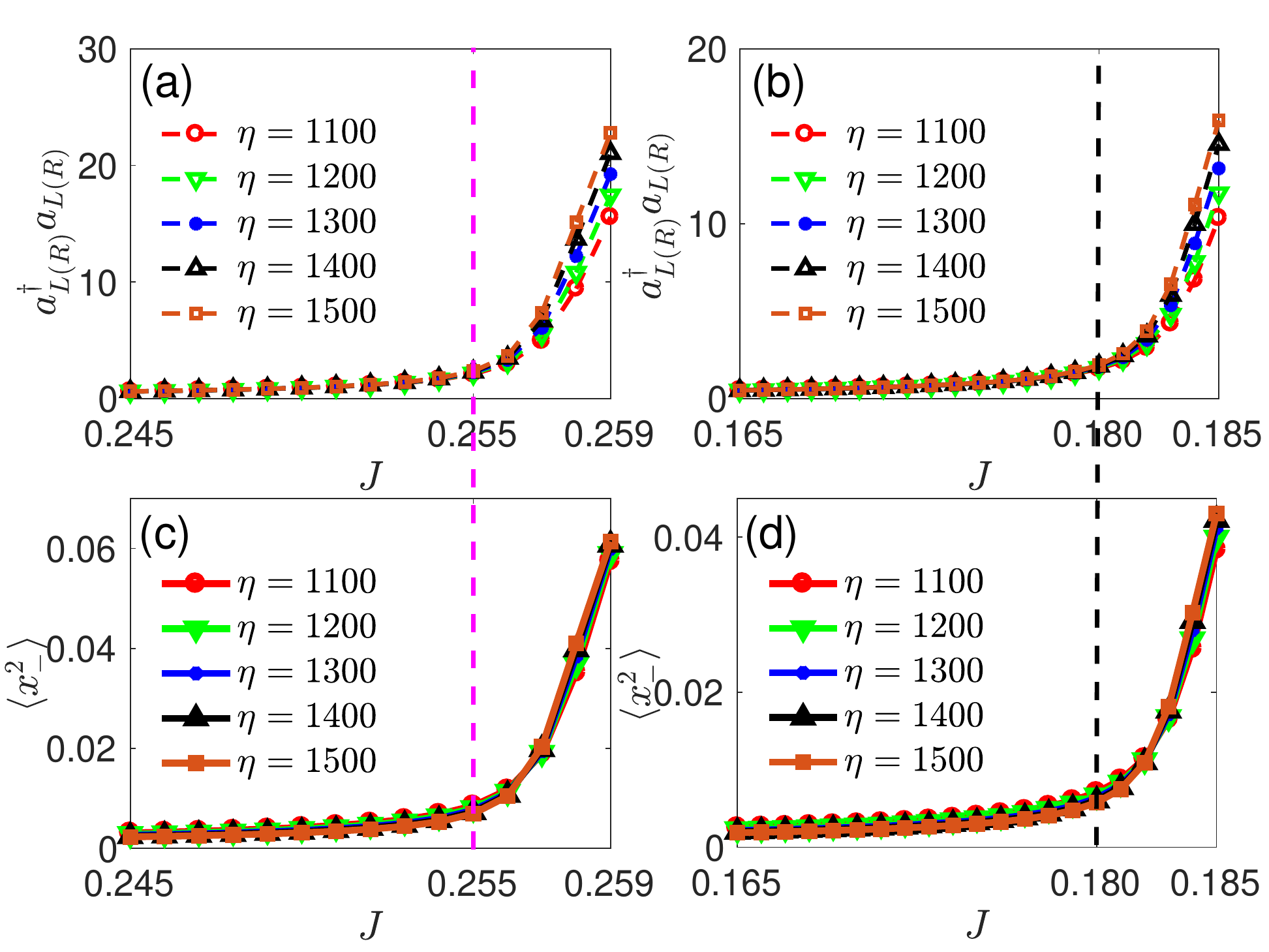}
		\caption{(Color Online) Upper panel: The GS photon populations $a^\dag_{L}a_{L}$ and $a^\dag_{R}a_{R}$
		as a function of the inter-cavity hopping strength $J$ with $g=0.7$ (a) and with $g=0.8$ (b).
		Lower panel: The expectation value of the squared anti-symmetric normal mode $\braket{x^2_{-}}$
		as a function of $J$ with $g=0.7$ (c) and with $g=0.8$ (d). The colors of the curves correspond to different $\eta$.
		The dotted magenta reference lines in (a) and (c) denote the exact phase boundaries for $g=0.7$. The
		dotted black reference lines in (b) and (d) denote the ones for $g=0.8$. }
		\label{f2}
\end{figure}

Then, we study the response of the GS photon populations $\braket{a^{\dag}_{L(R)}a_{L(R)}}$ to the
hopping strength $J$. By considering the above mentioned $\eta$ and the two cases where $g=0.7$
and $g=0.8$, we plot the $\braket{a^{\dag}_{L(R)}a_{L(R)}}$ as a function of $J$ in Figs.~\ref{f2}(a)
and \ref{f2}(b), respectively. Intuitively, there appears a significant feature that signals the occurrence
of the SPT. Namely, a large number of photons erupt in the cavities when the hopping strength $J$
exceeds the critical point \cite{R_d1}, before which the photon population approaches zero, corresponding
to the normal phase. Due to the parity symmetry, the photon populations of the two cavities vary synchronously
as the hopping strength $J$ increases. It can be interpreted why the photon population presents different
behavior before and after the SPT according to the change of the GS energy. When the system is in
the normal phase, the GS energy is stable at $E_{0}\approx-\eta$, without radiant photon. Therefore,
the GS photon populations of the two cavities both approach zero. When the system is in the superradiant
phase, due to the dramatic decrease of the GS energy with increasing $J$, macroscopic excitation of
photons emerges. Figures \ref{f2}(c) ($g=0.7$) and \ref{f2}(d) ($g=0.8$) present the change
of the GS expectation value of the squared anti-symmetric normal mode $\braket{x^2_{-}}$ with $J$.
This quantity serves as the order parameter to distinguish the superradiant phase from the normal phase.
Intuitively, the variational tendency is similar to that of the GS photon population. Before the SPT,
$\braket{x^2_{-}}$ approaches zero, whereas it tends to finite values when SPT happens. The superradiant 
phenomenon can be interpreted by the two-fold broken parity as well. On the one hand, the known shift 
of the oscillators to nonzero expectation values of $x^{2}_{-}$ leads to the nonzero $\langle a^\dag_{L(R)}a_{L(R)} \rangle$, 
but on the other, in the RD model, the breaking of the permutation system between cavities leads to the 
nonzero $\langle x^{2}_{-}\rangle$. In the above analyses, we have shown that the inter-cavity hopping parameter
in the RD model will lead to the SPT as well, which can be characterized by the behaviors of the GS energy
and the GS photon population, as well as the GS expectation value of the anti-symmetric normal mode.
In the following, we study the critical behavior and extract the CLCE by employing the fidelity susceptibility.

\section{Fidelity, Fidelity susceptibility  and quantum criticality} \label{S3}
\begin{figure}[htp]
		\centering
		\includegraphics[width=\textwidth]{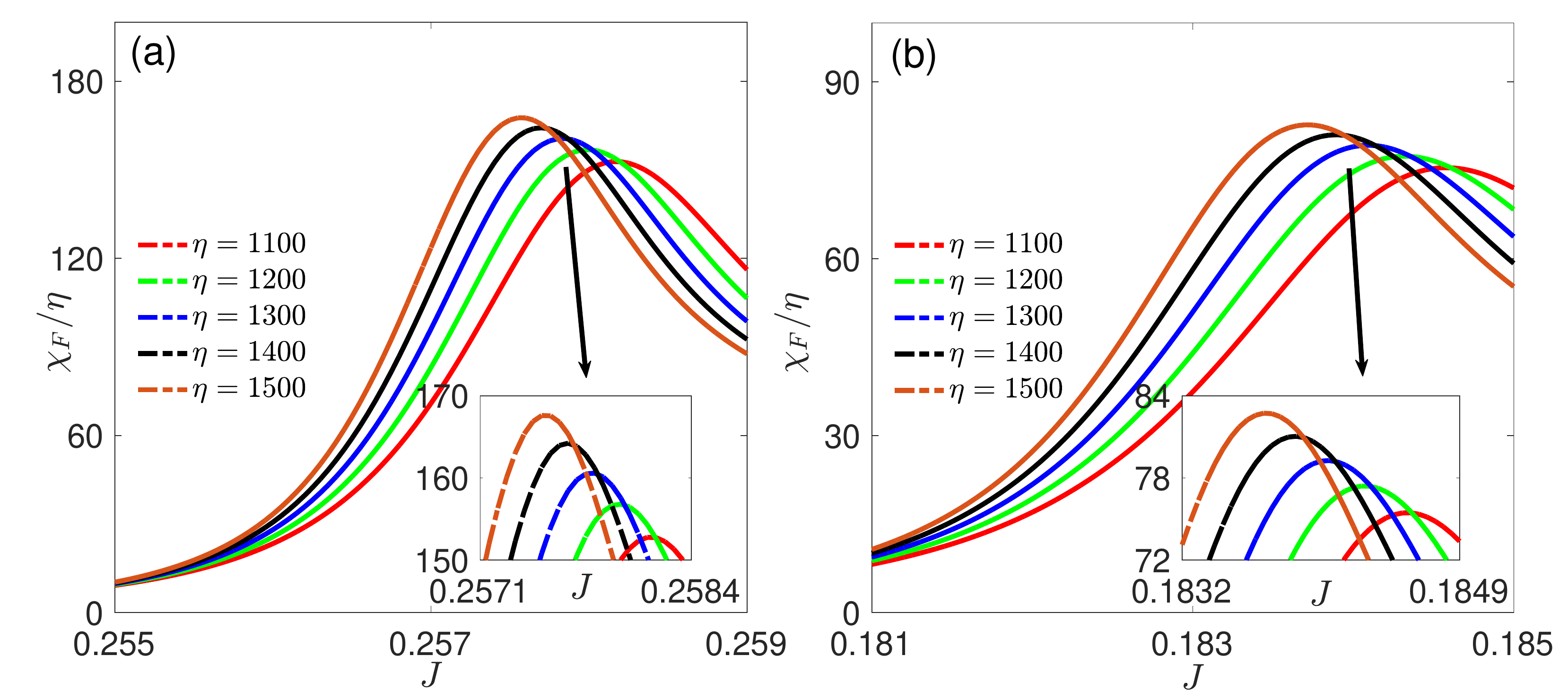}
		\caption{(Color online) The fidelity susceptibility $\chi_{F}$ (rescaled by $\eta$) as a function
		of the control parameter $J$. (a) $g=0.7$. (b) $g=0.8$. The colors of the curves correspond to different $\eta$.
		The inset represents the zoom around the peaks of $\chi_{F}/\eta$. Other involved parameter is $\delta{J}=5\times10^{-5}$. }
		\label{f3}
\end{figure}

The concept of fidelity originates from the quantum information theory \cite{qi} and is usually used to study the critical phenomenon.
The definition of the ground-state fidelity $F(J,\delta J)$ is presented as \cite{fidelity_1,fidelity_2,fidelity_3}
\begin {equation}
F(J,\delta J)=|\braket{\psi_{0}(J)|\psi_{0}(J+\delta J)}|,
\end{equation}
which is the overlap of two ground states $\psi_{0}(J)$ and $\psi_{0}(J+\delta J)$ with the driving parameter
having a small amount deviation $\delta J$. When $\delta J\rightarrow 0$, the
fidelity susceptibility (FS) is defined as \cite{FS_2}
\begin{equation}\label{XF_GS}
\chi_{F}=\lim_{\delta J\rightarrow 0}\frac{-2\ln F(J,\delta J)}{(\delta J)^2}.
\end{equation}

If considering the contribution of the excited states to the fidelity, the FS can be alternatively calculated by
$\chi_{F}(J)=\sum_{n\neq 0} \frac{|\braket{\psi_{n}(J)|H_{1}|\psi_{0}(J)}|^2}{\left[E_{n}(J)-E_{0}(J)\right]^2}$~\cite{FS_1,FS_2},
where the wave function $\psi_{n}(J)$ and eigenenergy $E_{n}$ satisfy $H(J)\ket{\psi_{n}(J)}=E_{n}(J)\ket{\psi_{n}(J)}$. This quantity is usually
employed to characterize the quantum phase transition and the quantum critical scaling behavior, and to extract the critical exponents \cite{qcce_1,qcce_2}.
If we calculate the FS according to the alternative definition, we need to ensure the convergence of the excited states. However, it requires a large truncation number and takes heavy computer time. Instead,  a large
truncation number can be avoided if we care only about the convergence of the ground state. Therefore, in our numerical simulation,
we employ Eq.~(\ref{XF_GS}) to compute the FS. The truncation number is taken at $N_{cut}=80$, which is sufficient for the
present purpose (See \ref{A} for the analysis of a larger truncation).

We concentrate on the critical behavior of the FS and extract the CLCE by performing the finite-frequency scaling analysis.
Referring to the definition of the finite-size scaling of the FS approaching the critical point that $\chi_{F}(J_{max}) \propto L^{\mu}$ \cite{fidelity_3,qcce_2,LMG_3},
the finite-frequency scaling of the FS around the transition point is given as
\begin{equation}\label{scaling_XF}
\chi_{F}(J_{max}) \propto \eta^{\mu},
\end{equation}
where $\mu$ is the critical exponent of FS and $J_{max}$ is the critical point at which the $\chi_{F}$ peaks.

For a finite-size system, the FS satisfies such a scaling behavior \cite{qcce_2,qcce_3}
\begin{equation}\label{XF}
\chi_{F}(J)=\eta^{2/\nu}f[\eta^{1/\nu}(J-J_{max})],
\end{equation}
where $\nu$ is the CLCE and $f(x)$ is the scaling function.

In the following, we study the quantum critical behavior. With the exact diagonalization method, we numerically obtain the full energy
spectrum and its corresponding wave function. By employing Eq.~(\ref{XF_GS}) and considering $\delta{J}=5\times10^{-5}$, we
plot the FS as a function of the control parameter $J$ for different $\eta$, which are plotted in Figs.~\ref{f3}(a) and \ref{f3}(b), respectively.
The corresponding $g$ are $g=0.7$ and $g=0.8$, respectively. The inset represents the zoom around the peaks of $\chi_{F}/\eta$.
The FS peaks at the critical point, showing the occurrence of the SPT. Besides, intuitively, one can see that the ratio $\eta$
plays a similar role as that of the size of the system in common phase transitions. The maximal value of FS varies with the
increasing $\eta$. Furthermore, the tendency of the $\chi_{F}/\eta$ is in accord with that of the single Rabi model \cite{qc_1},
namely, $\chi_{F}(J_{max})$ increases as the $\eta$ increases.

\begin{figure}[htp]
		\centering
		\includegraphics[width=\textwidth]{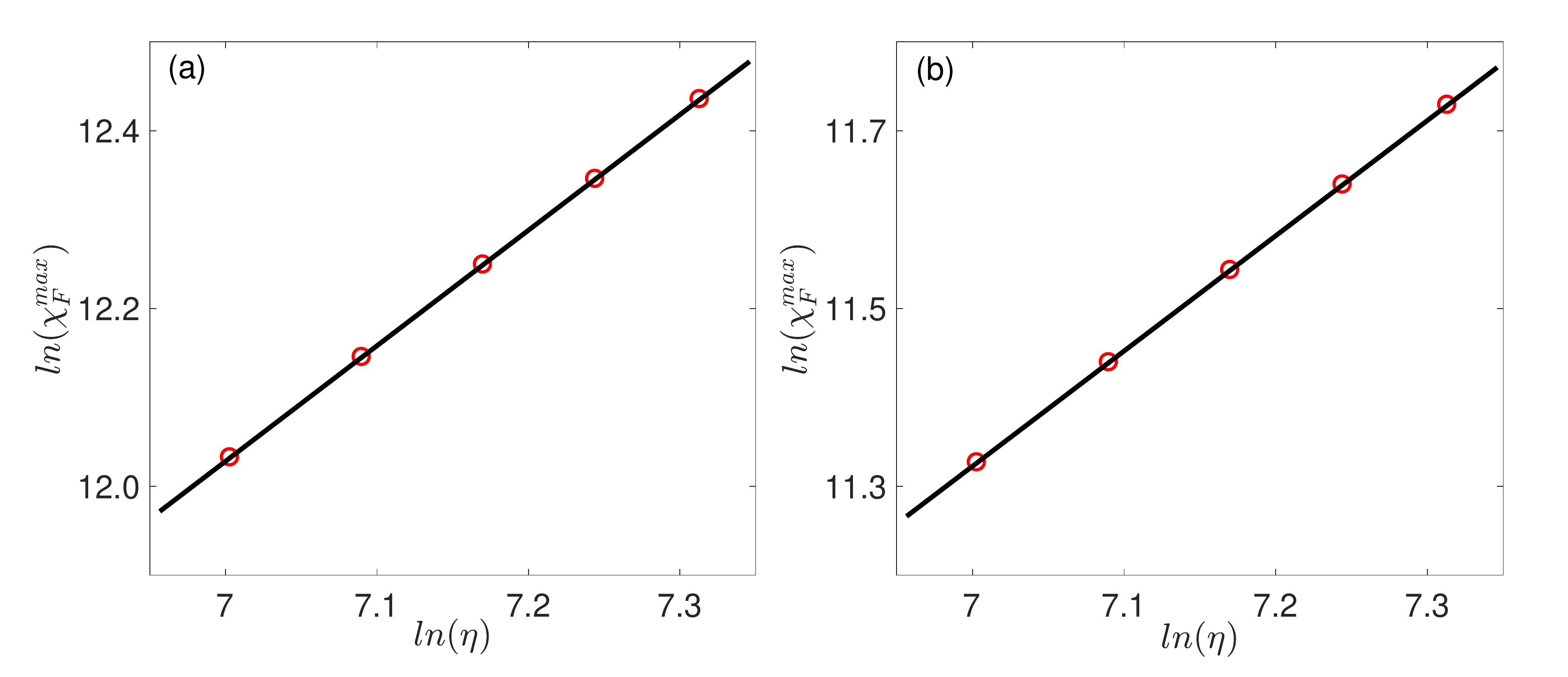}
		\caption{(Color online) Finite-frequency scaling of the logarithm of the maximal value of the
		fidelity susceptibility $ln(\chi_{F}^{max})$ as a function of $ln(\eta)$ for $g=0.7$ in (a) and for $g=0.8$ in (b).
		The red open circles are the numerical result and the black solid lines are the linear fitting of the
		numerically obtained data. In (a), the slope of the fitting curve presents the scaling exponent of the
		fidelity susceptibility $\mu=1.300\pm{0.001}$. In (b), the slope of the fitting curve presents the scaling
		exponent $\mu=1.297\pm{0.002}$. The involved parameter is $\delta{J}=10^{-5}$.}
		\label{f4}
\end{figure}

To verify the above conjecture, we first investigate the finite-frequency scaling of the FS near the critical point.
Figures \ref{f4}(a) and \ref{f4}(b) shows the logarithm of the maximal value of the fidelity susceptibility
$ln(\chi_{F}^{max})$ (the red open circles show) as a function of $ln(\eta)$, corresponding to the cases with $g=0.7$
and $g=0.8$, respectively. Here, $\delta=10^{-5}$ is considered in the calculations. After performing a linear fitting on
these data, we find that the slope of the fitting curve satisfies $1.300\pm{0.001}$ (the black line shows) for $g=0.7$.
From the scaling form in Eq.~(\ref{scaling_XF}), for $g=0.7$, the critical exponent of the FS near the critical point is
$\mu=2/\nu=1.300\pm{0.001}$, from which we extract the CLCE as $\nu|_{g=0.7}=1.539\pm{0.001}$. For $g=0.8$,
the slope of the fitting curve presents the exponent $\mu=2/\nu=1.297\pm{0.002}$. Therefore, the CLCE of the case
with $g=0.8$ is $\nu|_{g=0.8}=1.542\pm{0.002}$. The numerically obtained $\mu$ and $\nu$ are close to the exact
results \cite{HP,ps2}, equal to $4/3$ and $3/2$, respectively. Noting that there are precedents where the
numerical exponent deviates from the theoretical value \cite{qc_1,qc_3,LMG_3}. Our results imply that the inter-cavity-hopping-parameter
dominated SPT has a similar quantum scaling behavior as the SPT driven by the cavity-atom interaction in the single Rabi model.

\begin{figure}[htp]
		\centering
		\includegraphics[width=\textwidth]{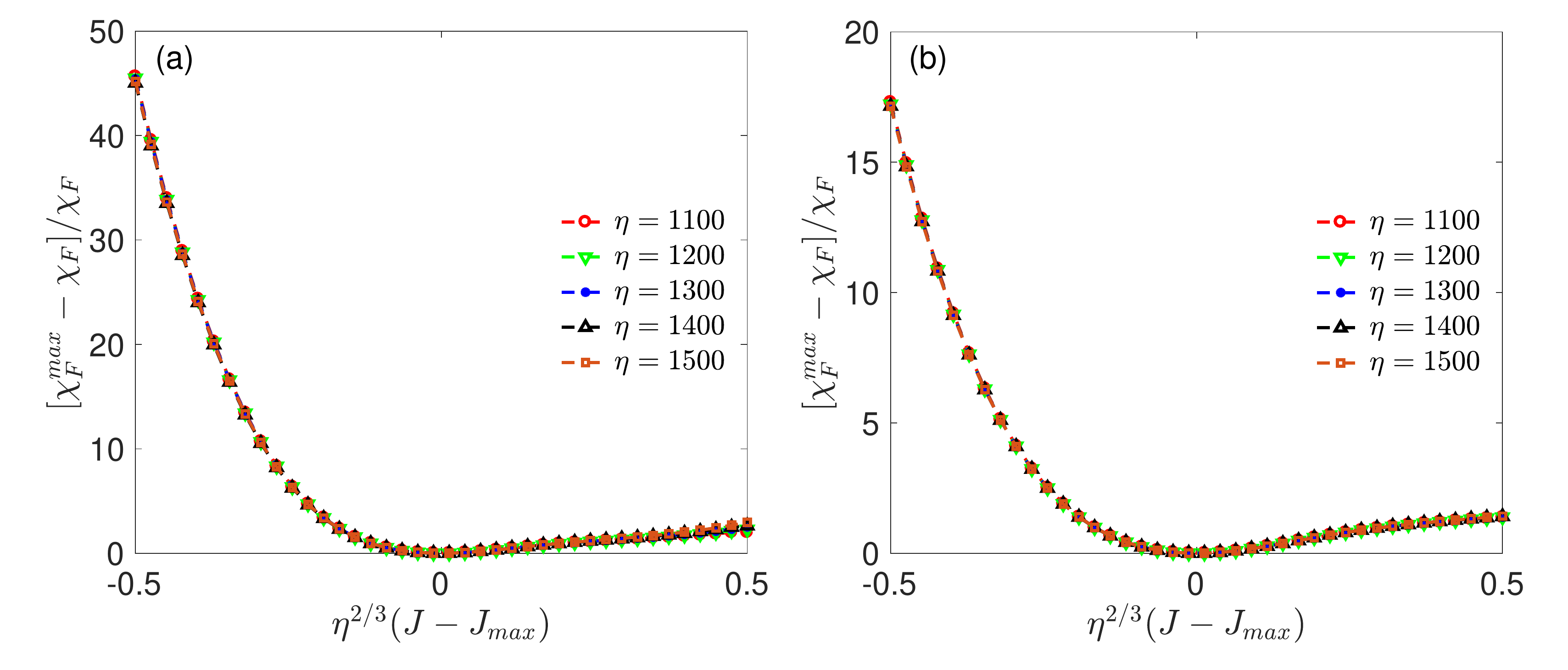}
		\caption{(Color online) Finite-frequency scaling of the rescaled FS according to Eq.~(\ref{CLCE}) with $g=0.7$ in (a)
		and with $g=0.8$ in (b). The colors of the curves correspond to different $\eta$. Other involved parameter is
		$\delta{J}=10^{-5}$.}
		\label{f5}
\end{figure}

We illustrate the quantum scaling behavior by the rescaled FS \cite{LMG_3,qcce_2}
\begin{equation}\label{CLCE}
\frac{\chi^{max}_{F}-\chi_{F}}{\chi_{F}}=f\left[\eta^{1/\nu}(J-J_{max})\right].
\end{equation}
The fraction $(\chi^{max}_{F}-\chi_{F})/\chi_{F}$ has reduced the leading coefficient $\eta^{2/\nu}$ in Eq.~(\ref{XF}),
making itself be a universal function of the rescaled control parameter $\eta^{1/\nu}(J-J_{max})$.
After taking $\delta{J}=10^{-5}$, the rescaled FS of cases $g=0.7$ and $g=0.8$ for different $\eta$ has been
presented in Fig.~\ref{f5}(a) and \ref{f5}(b), respectively. Intuitively, the rescaled FS  with various $\eta$ collapse
onto a single curve with CLCE $\nu=3/2$, signaling the same universal class in the single Rabi model, RD model,
Dicke model, and the LMG model.

\section{Summary}\label{S4}
In conclusion, an inter-cavity-hopping-parameter driven SPT in the Rabi-dimer model has been studied.
We have shown that the superradiance transition can be characterized by the GS energy and the
GS photon population, as well as the GS expectation value of the squared anti-symmetric normal mode.
In the normal phase, we find that the GS energy is invariant as the frequency ratio increases and
no photon population can be observed in the ground state. On the contrary, in the superradiant phase,
accompanied by the decrease of the GS energy, macroscopic photon excitation emerges.
Moreover, the trend of the GS expectation value of the anti-symmetric mode is similar to that of the GS photon population.
In the normal phase, the expectation value approaches zero, whereas it becomes a finite number when the system is in the
superradiant phase. After performing the scaling analysis on the fidelity susceptibility, we verified that such an SPT
belongs to the same universal class as the phase transition in the single Rabi model, RD model, Dicke model, and the LMG model.
Our study promotes a further understanding of the SPT in the Rabi-dimer model. We notice that recently there is a work on
the experimental realization of the Rabi-Hubbard model with trapped ions \cite{exp_Rabi_Hubbard}. The Rabi-dimer model can be
regarded as a minimal multi-cavity system of the Rabi-Hubbard model. We hope our theoretical finding that the quantum critical
behavior driven by the inter-cavity coupling can be confirmed by the quantum simulation with trapped ions.

\section*{ACKNOWLWDGMENTS}
We thank Maoxin Liu, Wen-long You, and Liwei Duan for their insightful discussions and acknowledge
Dr. Akhtar Naeem for his effort in polishing the English. This work is supported by the NSFC
under Grants No.~11835011 and No.~12174346.

\begin{appendix}
\section{Results of $g=0.5$}\label{A}
\begin{figure*}[htp]
		\centering
		\includegraphics[width=\textwidth]{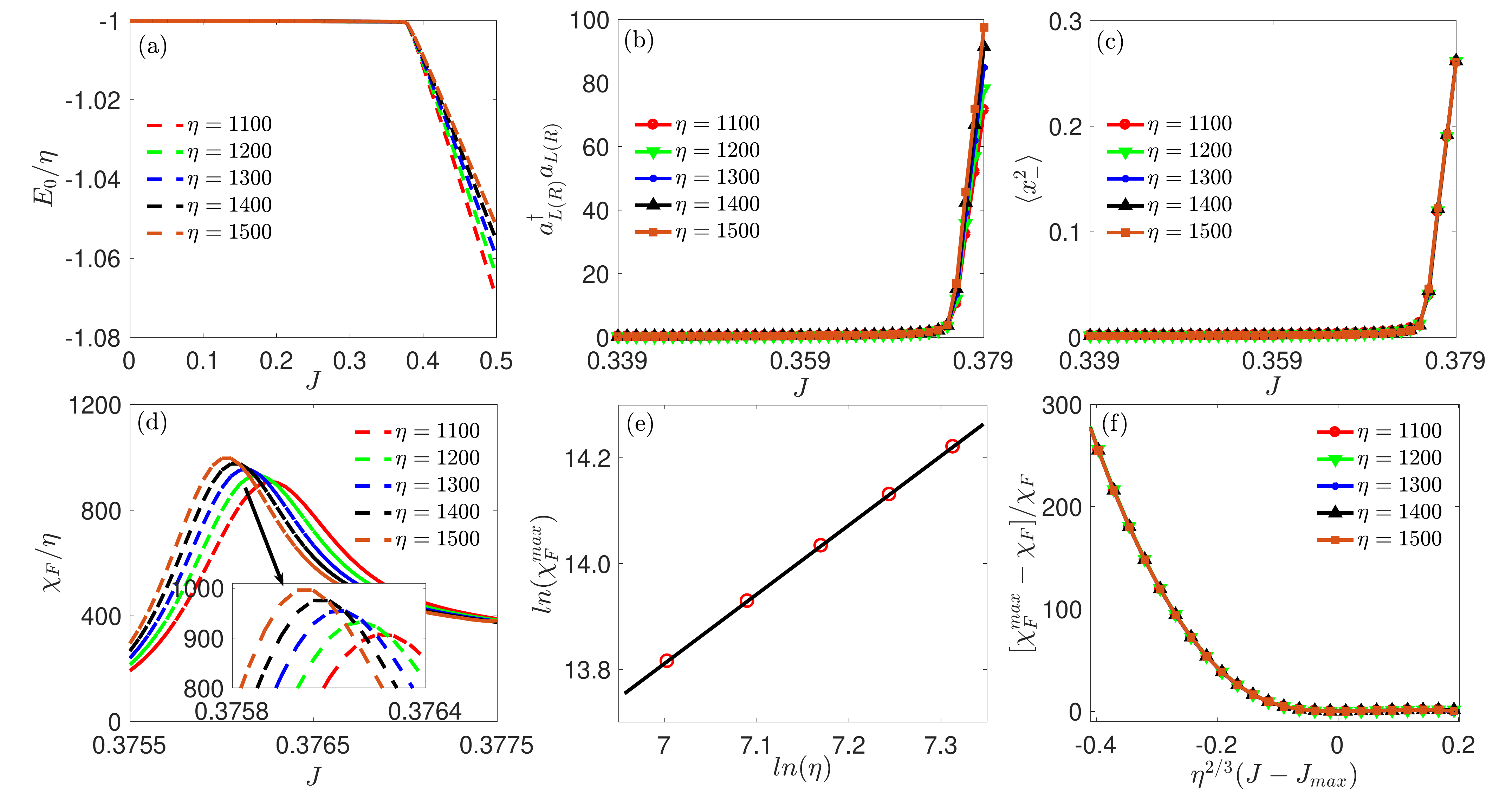}
		\caption{(Color online)
		(a) GS energy $E_{0}$ as a function of $J$ for different $\eta$.
		(b) GS photon population $\braket{a^{\dag}_{L(R)}a_{L(R)}}$ versus $J$ for different $\eta$.
		(c) GS expectation value of squared anti-symmetric normal mode $\langle x^2_{-}\rangle$
		varies with $J$ for different $\eta$.
		(d) The fidelity susceptibility $\chi_{F}$ versus $J$ for different $\eta$. The inset represents the zoom
		around the peak of $\chi_{F}/\eta$. The involved parameter is $\delta{J}=5\times10^{-5}$.
		(e) Finite-frequency scaling analysis of the maximal $\chi_{F}$ with $\delta{J}=10^{-5}$ (numerically
		obtained data are show with red circles). The slope of the fitting curve (black line) presents the scaling \
		exponent of the fidelity susceptibility $\mu=1.308\pm{0.001}$.
		(f) The rescaled FS as a function of $J$ for different $\eta$. The involved parameter is $\delta{J}=10^{-5}$.}
		\label{f6}
\end{figure*}

In this appendix, we present the numerical analysis about the case with $g=0.5$ and $N_{cut}=180$.
Figure \ref{f6}(a) plots the GS energy $E_{0}$ as a function of the cavity-coupling parameter $J$.
Intuitively, $E_{0}\approx -\eta$ before the SPT, and it tends to lower values when the system enters
into the superradiant phase. In analogy to the cases for $g=0.7$ and $g=0.8$, there are macroscopic
excitation photons accompanying the decrease of $E_{0}$, indicating the occurrence of SPT (see Fig.~\ref{f6}(b)).
The trend of the GS expectation value of the squared anti-symmetric normal mode $\langle x^2_{-}\rangle$
is similar to that of the GS photon population. As the Fig.~\ref{f6}(c) shows, when the system is in the
normal phase, $\langle x^{2}_{-}\rangle\approx 0$, whereas it goes to finite values when $J$ crosses
the critical point. To investigate the quantum criticality of this case, we calculate the FS according to
Eq.~(\ref{XF_GS}) for various $\eta$, which are shown in Fig.~\ref{f6}(d) (where the inset represents
the zoom around the peak of $\chi_{F}/\eta$). We can see that, when $\eta$ gets larger, the peak
of $\chi_{F}/\eta$ gets larger and the corresponding $J_{max}$ is closer to the theoretical value
$J^{g=0.5}_{c}=0.375$. The data fitting of $\chi^{max}_{F}$ is presented in Fig.~\ref{f6}(e). The red
circles denote the numerically obtained data. The black solid line is the fitting curve of these data,
whose slope presents the scaling exponent of FS $\mu=1.308\pm0.001$. Furthermore, we calculate the
CLCE by $\mu=2/\nu$, and obtain the CLCE is $\nu=1.529\pm 0.001$, which is closer to
the theoretical value $\nu=3/2$ than those of the $N_{cut}=80$ cases. Although there is a slight deviation
between the obtained $\nu$ and the theoretical value $\nu=3/2$, the rescaled FS plotted in Fig.~\ref{f6}(f)
with various $\eta$ show that they all collapse onto a single curve with CLCE $\nu=3/2$.
\end{appendix}



%% References


\bibliographystyle{elsarticle-num}
\bibliography{reference}






% \end{thebibliography}

\end{document}